\documentclass[conference]{IEEEtran}

\usepackage{amsmath, amssymb}
\usepackage{graphicx}
\usepackage{booktabs}
\usepackage{multirow}
\usepackage{array}
\usepackage{cite}
\usepackage{hyperref}
\usepackage{xcolor}
\usepackage{setspace}
\usepackage{tikz}
\usepackage{pgfplots}
\pgfplotsset{compat=1.17}
\usepackage{placeins}

\usepackage{capt-of}
\usepackage{etoolbox}
\usepackage[numbers,  sort&compress]{natbib}
\usepackage[labelfont=bf,labelsep=colon]{caption}
\usepackage[
    letterpaper,
    top=0.75in,
    bottom=0.973in,
    left=0.6in,
    right=0.6in
]{geometry}
\title{WINT: A Novel Weighted Integer Representation with Improved Error Characteristics}

\author{
\IEEEauthorblockN{ Cheng-Yen Lee, Zach Assad, Gautham Nemani, Sunil P. Khatri}
\IEEEauthorblockA{
Department of Electrical and Computer Engineering, Texas A\&M University, College Station, TX, USA\\
\{cylee, xachariah, gnemani, sunilkhatri\}@tamu.edu
}
\vspace{-3em}
}

\begin{document}
\makeatletter
\patchcmd{\@maketitle}{\vskip 1.0em}{\vskip 0.1em}{}{}
\makeatother

\maketitle

\begin{abstract}
In computing, there is a need for number representation schemes that provide large dynamic range with low error. Many applications, including embedded systems and edge machine learning, have stringent memory constraints yet require large dynamic range for data representation. We present Weighted Integer (WINT), a simple and configurable mantissa-exponent number format with user-selectable mantissa ($m$) and exponent ($e$) bit allocations (also referred to as configurations) that enables application-specific precision versus range tradeoffs at design time. We develop a complete analytical framework for computing Mean Relative Error (MRE), the primary metric for characterizing WINT's error. Since exact MRE calculations grow exponentially with mantissa size, we introduce harmonic and Taylor series approximation methods that achieve O(1) time complexity regardless of configuration. The Taylor series and harmonic approximations demonstrate significant speedups over the exact method while maintaining accuracy within 0.2\% for the configurations presented. Our experiments across 8 to 32-bit configurations show that allocating 2 exponent bits consistently yields both lower MRE by 12--33\% and $2\times$ greater range than the integer baseline for bit widths of 12 and above. Allocating 3 exponent bits extends range by $16\times$ while reducing MRE by 15--50\% for bit widths of 16 and above.
\end{abstract}


\vspace{-3mm}
\section{Introduction}
\label{sec:introduction}

Number representation is fundamental to digital computing. The choice of representation determines what values can be stored, how much memory is required, and what errors may be incurred. These factors directly affect both the correctness and efficiency of computation, with implications that vary widely across application domains. A format well-suited for scientific simulation may be unnecessarily complex for an embedded controller, while a format optimized for neural network inference may lack the precision required for financial calculations. No single format is optimal for all applications; rather, users must select representations that best match their specific requirements for range, precision, and implementation complexity, often under tight memory and power constraints. Two dominant representations are integer and floating-point, each with distinct trade-offs.

Standard unsigned integers represent whole numbers from 0 to $2^K - 1$, where $K$ is the total bit width. Integer representation is simple, widely supported, and requires no decoding beyond reading the binary value directly. However, the range is limited by the bit width---a 16-bit integer cannot represent values larger than 65,535. Furthermore, when quantizing real non-integer values to integers, rounding to the nearest whole number incurs representation error.

Floating-point formats, such as IEEE 754~\cite{ieee754_citation}, overcome the range and precision limitations of integers by partitioning the total $K$ bits into three fields: a 1-bit sign value $S$, an $m$-bit mantissa $M$, and an $e$-bit exponent $E$, such that $K = 1 + m + e$. The value is computed as $(-1)^S \times (1.M) \times 2^{(E-\text{Bias})}$, where the implicit leading 1 and exponent ``Bias" are defined by the standard. However, this expanded range and precision comes at the cost of complexity: decoding requires bias subtraction, handling of an implicit leading bit, and management of special encodings for NaN and infinity. The bit allocations are also fixed by the standard---FP32, for example, mandates 8 exponent bits and 23 mantissa bits, leaving no room for application-specific tuning.

In many applications, memory capacity is a critical constraint that restricts the bit width for data representation. For example, embedded systems and sensors operate under strict storage limits~\cite{embedded_quantization}, while machine learning models with millions of parameters benefit significantly from reduced precision formats with large dynamic ranges that lower memory footprint without unacceptable loss of accuracy~\cite{fp8_citation}. For such memory-constrained applications, a number representation whose range exceeds that of integers while remaining simpler to realize than full floating-point is highly desirable.

In this paper, we present Weighted Integer (WINT), a configurable number format that bridges the gap between limited-range integers and complex floating-point formats with a larger range and better error characteristics than integers. WINT uses a mantissa-exponent structure like floating-point, but without the complexity of implicit bits, bias values, or special encodings. The value of a WINT number is simply $M \times 2^E$, where both $M$ and $E$ are unsigned integers. The bit allocation between mantissa ($m$ bits) and exponent ($e$ bits) is user-selectable, enabling application-specific precision versus range trade-offs. Compared to floating-point, WINT offers simpler encoding and decoding while allowing flexible bit allocation. Compared to standard integers, WINT provides larger range for the same bit width and better error characteristics. We adopt the Mean Relative Error (MRE) as the primary metric for evaluating error. MRE measures the relative error between an actual value being represented and its nearest representable WINT value. Computing MRE exactly takes exponentially longer as the bit width of the mantissa increases. By deriving approximate closed-form expressions for MRE which have an O(1) time complexity, we achieve significant computation speedups over the exact calculation, with minimal accuracy loss.

The contributions of this paper are as follows:
\begin{itemize}
\item We present the Weighted Integer (WINT) number representation, as an alternative to the standard integer representation.
\item We develop a complete analytical framework for computing Mean Relative Error (MRE) to compare the error characteristics of WINT with standard integers.
\item We introduce novel harmonic and Taylor series approximation methods for computing MRE that achieve O(1) time complexity and similar accuracy (within $0.2\%$) compared to the exact MRE computation.
\item We demonstrate that WINT can effectively trade off MRE and range, and for many configurations, improve over standard integers in both metrics. For example, for a 32-bit WINT configuration, we can achieve a range of $6.87 \times 10^{10}$ with MRE of $7.24 \times 10^{-10}$ compared to a 32-bit integer which has a range of $4.29 \times 10^{9}$ and MRE of $1.45 \times 10^{-9}$.
\end{itemize}

The remainder of this paper is organized as follows. Section~\ref{sec:related_work} reviews related work. Section~\ref{sec:wint_format} defines the WINT format. Section~\ref{sec:error_analysis_approach} presents the error analysis methodology. Section~\ref{sec:results} provides experimental results. Section~\ref{sec:conclusion} concludes our paper.

\section{Related Work}
\label{sec:related_work}

This section reviews existing number representation schemes. We organize related work into three categories: block-based formats, scalar floating-point formats, and alternative encodings.

\subsection{Block-Based Formats}
\label{subsec:block_based}

Block-based formats represent a group (or block) of numbers that share a common scaling factor, amortizing the cost of storing exponent information across multiple numbers. Given a block of $k$ numbers, the value $V_i$ of each number is generally computed as:
\begin{equation}
V_i = M_i \times 2^{E_{\text{shared}}}
\label{eq:block_format}
\end{equation}
where $M_i$ is the individual mantissa and $E_{\text{shared}}$ is the exponent common to all elements in the block. This approach reduces storage compared to formats where each number has its own exponent, but requires that all numbers in a block share the same exponent.

Block Floating Point (BFP)~\cite{bfp_citation} uses traditional floating-point mantissas with a shared exponent per block, where the shared exponent is typically determined by the largest magnitude element in the block. Microscaling (MX) formats~\cite{mx_citation} standardize block-based representations for machine learning, supporting both integer and low-precision floating-point element types within blocks; however, the floating-point elements are limited to a small set of fixed configurations (e.g., FP8, FP6, FP4) with predetermined mantissa and exponent bit allocations. Flexpoint~\cite{flexpoint_citation} uses fixed-point mantissas with a shared exponent per block, where the shared exponent is automatically adjusted during neural network training based on statistical analysis of block values. Flexpoint also supports overflow prediction by tracking the probability that computations will exceed the representable range, allowing preemptive exponent adjustment.

WINT differs from block-based formats in that each number is independently representable. Block-based formats require multiple values to share a common exponent, constraining the dynamic range of individual elements to the exponent of the block. In contrast, a WINT number is decoded using only its own contiguous mantissa and exponent bit fields, without requiring access to any shared exponent information. Additionally, WINT requires no runtime exponent management or overflow prediction---the bit allocation is fixed at design time, and the representable range is statically determined at the time of choosing the configuration.

WINT is a generalization of BFP with integer mantissas, with 2 key differences. First, WINT includes the exponent in the number, without needing a shared exponent. This increases representational flexibility. Second, the size of both the mantissa and exponent \{m, e\} are both flexible for WINT, and the user can choose them according to application requirements.

To the extent that any of the block-based formats use integer mantissas and unbiased integer exponents, they can benefit from the error analysis presented in this paper (such as MX with integer mantissas).

\vspace{-3mm}

\subsection{Scalar Floating-Point Formats}
\label{subsec:scalar_fp}

Scalar floating-point formats represent each number independently by partitioning the available bits into three fields: a 1-bit sign $S$, an $m$-bit mantissa $M$, and an $e$-bit exponent $E$. IEEE 754~\cite{ieee754_citation} defines standard formats including FP16, FP32, and FP64. Values are computed as:
\begin{equation}
V = (-1)^S \times (1.M) \times 2^{(E-\text{$E_0$})}
\label{eq:ieee_format}
\end{equation}
where the mantissa is fractional, and has an implicit leading 1, and the exponent is offset by a bias $E_0$ defined by the standard. Special bit patterns are reserved for NaN and infinity. In addition to IEEE 754 standard formats, recent low-precision variants such as FP8~\cite{fp8_citation}, FP6, and FP4~\cite{mx_citation, ocp_mx_spec} target machine learning applications.

Posit~\cite{posit_citation} is an alternative scalar format that uses variable-length fields: a 1-bit sign $S$, a variable-length regime field, an exponent $E$ (up to $es$ bits), and a fraction $F$. The regime field consists of a run of identical bits terminated by an opposite bit, encoding a run-length value $k$ that determines a large scaling factor. Values are computed as:
\begin{equation}
V = (-1)^S \times \textit{useed}^k \times 2^{E} \times (1+F)
\label{eq:posit_format}
\end{equation}
where $\textit{useed} = 2^{2^{es}}$ and $k$ is the run-length of the regime bits. Unlike IEEE 754 formats with fixed field widths, Posit's regime field expands or contracts based on the value's magnitude, which in turn compresses or expands the remaining bits available for the exponent and fraction. This variable-length encoding provides \emph{tapered} precision: values near 1 have short regime fields, leaving more bits for the fraction and thus higher precision, while values far from 1 require longer regime fields, reducing precision.

WINT differs from these formats in several ways. First, the value of any WINT number is always integral. Secondly, the bit allocation between mantissa and exponent \{m, e\} is fully user-selectable, allowing users to choose the configuration that best suits their application's requirements. Third, WINT uses no implicit leading bit, bias, or special value encodings; all $2^{m+e}$ bit patterns represent numerical values, and decoding is a simple operation.

Although WINT, as presented in this paper, only represents non-negative numbers, this can easily be changed by adding a sign bit, or using a 2's complement mantissa field. 

\subsection{Alternative Encodings}
\label{subsec:alternative}

Beyond block-based and scalar floating-point formats, several alternative encodings adopt fundamentally different strategies for representing numerical values. These approaches do not fit neatly into the previous categories and offer distinct computational trade-offs.

The Logarithmic Number System (LNS)~\cite{number_systems_survey} represents a value by storing the logarithm of its magnitude. Given a value $V$, LNS stores $\log_a(|V|)$ as a fixed-point number with $a$ being the radix, plus a sign bit. This simplifies multiplication and division (which become addition and subtraction in the log domain) but complicates addition and subtraction, requiring lookup tables or iterative approximations.

The Residue Number System (RNS)~\cite{number_systems_survey} represents a value as a vector of residues modulo a set of coprime integers. Arithmetic operations proceed independently in each residue channel, enabling carry-free parallel computation. However, comparison, division, and overflow detection are difficult to perform in RNS and require costly conversions~\cite{rns_isupov}, and the moduli must be carefully selected.

WINT avoids the domain conversions required by LNS and the modular arithmetic of RNS. Numbers are represented in a straightforward mantissa-exponent form, requiring no logarithmic transformations or residue decomposition. Additionally, WINT imposes no constraints on bit allocation, unlike RNS where system correctness depends on appropriate modulus selection.

\vspace{-4mm}

\section{WINT}
\label{sec:wint_format}

In this section, we present the Weighted Integer (WINT) format. We first define the format in Section~\ref{subsec:wint_definition}, then describe its fundamental properties in Section~\ref{subsec:wint_properties}, and introduce the error metric (MRE) used to characterize its error in Section~\ref{subsec:error_metric_introduction}.
\vspace{-1mm}
\subsection{WINT Definition}
\label{subsec:wint_definition}

We define the Weighted Integer format, illustrating its properties using a $6$-bit example configuration with $\{m=4, e=2\}$ as a running example.

WINT represents a number using two unsigned integer fields: an $m$-bit mantissa $M$ and an $e$-bit exponent $E$. The value of the WINT number is computed as:
\begin{equation}
V = M \times 2^E
\label{eq:wint_value}
\end{equation}
where $M$ ranges from $0$ to $2^m - 1$ and $E$ ranges from $0$ to $2^e - 1$. The total bit width is $K = m + e$.

\textbf{Example 1:} For a $6$-bit WINT with $\{m=4, e=2\}$, suppose $M = 7$ and $E = 2$. Then the value is $V = 7 \times 2^2 = 28$.

Note that WINT can represent certain values in multiple ways. For instance, the value $28$ can also be represented with $M = 14$ and $E = 1$ (since $14 \times 2^1 = 28$). This means the WINT encoding is non-canonical, and the number of unique representable values is less than $2^{m+e}$. A detailed analysis of the unique value count and its derivation is presented in Section~\ref{sec:error_analysis_approach}.

The WINT format as analyzed in this paper is unsigned, representing non-negative values. A signed extension is straightforward: adding a 1-bit sign value $S$ gives $V = (-1)^S \times M \times 2^E$. Alternatively, one could use 2's complement to represent the mantissa. The error characteristics remain identical since the magnitude distribution is unchanged. This paper focuses on unsigned WINT; the signed case follows directly.

\subsection{WINT Properties}
\label{subsec:wint_properties}

This subsection describes the fundamental precision versus range trade-off inherent to the WINT format. The user's choice of mantissa ($m$) and exponent ($e$) bit allocation determines both the maximum representable range and the representation error, making this trade-off the central consideration when selecting a WINT configuration.

For a fixed total bit width $K$, WINT can provide larger range than a standard $K$-bit unsigned integer, at the cost of reduced precision. The maximum representable value is:
\begin{equation}
V_{\max} = (2^m - 1) \times 2^{(2^e - 1)}
\label{eq:wint_max}
\end{equation}

\textbf{Example 2:} For a $6$-bit WINT with $\{m=4, e=2\}$, the maximum mantissa is $2^{4} - 1 = 15$ and the maximum exponent is $2^2 - 1 = 3$, giving $V_{\max} = 15 \times 2^{3} = 120$. By comparison, a $6$-bit unsigned integer has a maximum of $2^6 - 1 = 63$, so WINT provides a range increase factor of $120 / 63 \approx 1.9\times$.

The range increase becomes more dramatic with larger bit widths. For a $16$-bit WINT with $\{m=12, e=4\}$, $V_{\max} = 4095 \times 2^{15} = 134{,}184{,}960$, compared to $65{,}535$ for a standard $16$-bit unsigned integer---an increase of over $2{,}000\times$.

The choice of $\{m, e\}$ determines the precision-range trade-off: more mantissa bits (larger $m$) yield finer precision but a smaller range, while more exponent bits (larger $e$) provide a larger range at the cost of coarser precision. The user selects $\{m, e\}$ at design time based on application requirements.

Interestingly, although more exponent bits yield lower precision, they frequently offer much better error characteristics which we explore next.
\vspace{-2mm}
\subsection{Error Metric Introduction}
\label{subsec:error_metric_introduction}

This subsection introduces the error metrics used throughout the remainder of the paper.

The \textit{representation error} for a true value is the distance between the \emph{true value} and the nearest \emph{representable WINT value}. We quantify this using \textit{relative error} (RE):
\begin{equation}
\text{RE} = \frac{|V_{\text{repr}} - V_{\text{true}}|}{V_{\text{true}}}
\label{eq:relative_error}
\end{equation}
where $V_{\text{true}}$ is the true (actual) value which we would like to represent, and $V_{\text{repr}}$ is the nearest representable WINT value.

We use relative error rather than absolute error because it is the relevant error metric at the application level.

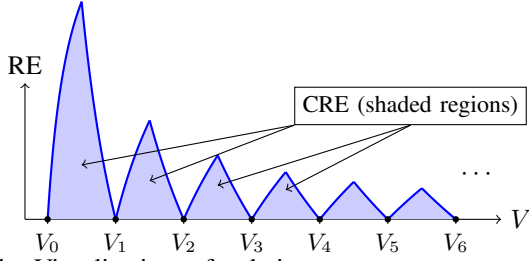
\begin{figure}[t]
    \centering
    \begin{tikzpicture}[scale=0.60]
        \draw[->] (0,0) -- (10.5,0) node[right] {$V$};
        \draw[->] (0,0) -- (0,3) node[above] {RE};
        
        \foreach \x/\label in {0.5/$V_0$, 2/$V_1$, 3.5/$V_2$, 5/$V_3$, 6.5/$V_4$, 8/$V_5$, 9.5/$V_6$} {
            \draw (\x,0) -- (\x,-0.1);
            \node[below] at (\x,-0.1) {\small \label};
        }
        
        \fill[blue!20] (0.5,0) -- plot[domain=0.5:1.25,samples=30] (\x,{8*(\x-0.5)/\x}) 
                       -- plot[domain=1.25:2,samples=30] (\x,{8*(2-\x)/\x}) -- (2,0) -- cycle;
        \draw[blue,thick] plot[domain=0.5:1.25,samples=30] (\x,{8*(\x-0.5)/\x});
        \draw[blue,thick] plot[domain=1.25:2,samples=30] (\x,{8*(2-\x)/\x});
        
        \fill[blue!20] (2,0) -- plot[domain=2:2.75,samples=30] (\x,{8*(\x-2)/\x}) 
                       -- plot[domain=2.75:3.5,samples=30] (\x,{8*(3.5-\x)/\x}) -- (3.5,0) -- cycle;
        \draw[blue,thick] plot[domain=2:2.75,samples=30] (\x,{8*(\x-2)/\x});
        \draw[blue,thick] plot[domain=2.75:3.5,samples=30] (\x,{8*(3.5-\x)/\x});
        
        \fill[blue!20] (3.5,0) -- plot[domain=3.5:4.25,samples=30] (\x,{8*(\x-3.5)/\x}) 
                       -- plot[domain=4.25:5,samples=30] (\x,{8*(5-\x)/\x}) -- (5,0) -- cycle;
        \draw[blue,thick] plot[domain=3.5:4.25,samples=30] (\x,{8*(\x-3.5)/\x});
        \draw[blue,thick] plot[domain=4.25:5,samples=30] (\x,{8*(5-\x)/\x});
        
        \fill[blue!20] (5,0) -- plot[domain=5:5.75,samples=30] (\x,{8*(\x-5)/\x}) 
                       -- plot[domain=5.75:6.5,samples=30] (\x,{8*(6.5-\x)/\x}) -- (6.5,0) -- cycle;
        \draw[blue,thick] plot[domain=5:5.75,samples=30] (\x,{8*(\x-5)/\x});
        \draw[blue,thick] plot[domain=5.75:6.5,samples=30] (\x,{8*(6.5-\x)/\x});
        
        \fill[blue!20] (6.5,0) -- plot[domain=6.5:7.25,samples=30] (\x,{8*(\x-6.5)/\x}) 
                       -- plot[domain=7.25:8,samples=30] (\x,{8*(8-\x)/\x}) -- (8,0) -- cycle;
        \draw[blue,thick] plot[domain=6.5:7.25,samples=30] (\x,{8*(\x-6.5)/\x});
        \draw[blue,thick] plot[domain=7.25:8,samples=30] (\x,{8*(8-\x)/\x});
        
        \fill[blue!20] (8,0) -- plot[domain=8:8.75,samples=30] (\x,{8*(\x-8)/\x}) 
                       -- plot[domain=8.75:9.5,samples=30] (\x,{8*(9.5-\x)/\x}) -- (9.5,0) -- cycle;
        \draw[blue,thick] plot[domain=8:8.75,samples=30] (\x,{8*(\x-8)/\x});
        \draw[blue,thick] plot[domain=8.75:9.5,samples=30] (\x,{8*(9.5-\x)/\x});
        
        \foreach \x in {0.5, 2, 3.5, 5, 6.5, 8, 9.5} {
            \fill (\x,0) circle (2pt);
        }
        
        \node[draw, fill=white, inner sep=3pt] (legend) at (8.5, 2.5) {\small CRE (shaded regions)};
        \draw[->] (legend.south west) -- (1.25, 1.2);
        \draw[->] (legend.south west) -- (2.75, 0.85);
        \draw[->] (legend.south) -- (4.25, 0.75);
        \draw[->] (legend.south) -- (5.75, 0.65);
        
        \node at (10, 1) {$\cdots$};
    \end{tikzpicture}
    \vspace{-3mm}
    \caption{Visualization of relative error across representable intervals. Within each interval $[V_k, V_{k+1})$, the relative error peaks at the midpoint $V_{\text{half}} = \frac{V_k + V_{k+1}}{2}$ and decreases to zero at the endpoints. The shaded areas represent the cumulative relative error (CRE) for each interval.}
    \label{fig:mre_visualization}
\end{figure}

The \textit{cumulative relative error} (CRE) for an interval between two consecutive representable values $[V_k, V_{k+1})$ is the total relative error accumulated across all possible true values within that interval. Figure~\ref{fig:mre_visualization} illustrates the relative error across multiple intervals.

The \textit{Mean Relative Error} (MRE) averages the CRE across all intervals between consecutive representable WINT values, providing a single metric to characterize the overall error characteristics of a number representation scheme.

The choice of $\{m, e\}$ influences both the representable range and the resulting MRE. Section~\ref{sec:results} presents comprehensive MRE measurements across WINT configurations from $8$ to $32$ bits.

Computing the exact MRE requires integration over all intervals between consecutive representable values---a computationally expensive process. Section~\ref{sec:error_analysis_approach} presents efficient methods for speeding up this computation with minimal accuracy loss.
\vspace{-2mm}
\section{Error Analysis Approach}
\label{sec:error_analysis_approach}

In this section, we derive methods for efficiently computing the Mean Relative Error (MRE) of WINT configurations. We begin with the ``Integration Method" (IM) (Section~\ref{subsec:integral_method}), which computes MRE exactly through direct integration over all representable intervals. We then develop the ``Triangle Approximation" (TA) (Section~\ref{subsec:triangle_method}), which simplifies computation by approximating the error profile geometrically. Finally, we present the Taylor series (TSA) (Section~\ref{subsec:taylor_approx}) and harmonic approximation (HA) (Section~\ref{subsec:harmonic_approx}) methods that achieve O(1) time complexity with minimal accuracy loss, enabling analysis of large configurations where exact computation is infeasible.

\subsection{Integration Method (IM)}
\label{subsec:integral_method}

For all methods, we treat true values as continuous rather than discrete, enabling precise characterization through integration. We define an \textit{interval} as the range between two consecutive representable WINT values $[V_k, V_{k+1})$. Any true value within this interval must be rounded to one of these endpoints, which are representable values. The \textit{cumulative relative error} (CRE) for an interval $[V_k, V_{k+1})$ is the total relative error accumulated across all possible true values within that interval. The \textit{Mean Relative Error} (MRE) is then the average CRE across all intervals in the representable range.

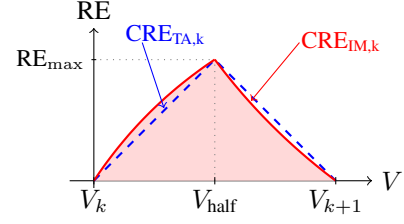
\begin{figure}[t]
\centering
\begin{tikzpicture}[scale=0.80, xscale=4, yscale=6]
\draw[->] (0.9,0) -- (2.15,0) node[right] {$V$};
\draw[->] (1,-.02) -- (1,0.42) node[above] {RE};

\fill[red!15] plot[smooth, domain=1:1.5, samples=50]
    (\x, {1 - 1/\x})
    -- plot[smooth, domain=1.5:2, samples=50]
    (\x, {2/\x - 1})
    -- (2,0) -- (1,0) -- cycle;

\draw[dashed, thick, blue] (1,0) -- (1.5,0.333) -- (2,0);

\draw[thick, red] plot[smooth, domain=1:1.5, samples=50]
    (\x, {1 - 1/\x});

\draw[thick, red] plot[smooth, domain=1.5:2, samples=50]
    (\x, {2/\x - 1});

\draw[dotted, gray] (1.5,0) -- (1.5,0.333);

\draw[dotted, gray] (1,0.333) -- (1.5,0.333);

\node[below] at (1,0) {$V_k$};
\node[below] at (1.5,0) {$V_{\text{half}}$};
\node[below] at (2,0) {$V_{k+1}$};

\node[left] at (1,0.333) {\small $\text{RE}_{\max}$};

\draw[->, red] (1.85,0.38) -- (1.65,0.22);
\node[red, right] at (1.82,0.38) {\small $\text{CRE}_{\text{IM,k}}$};

\draw[->, blue] (1.2,0.38) -- (1.29,0.2);
\node[blue, left] at (1.5,0.4) {\small $\text{CRE}_{\text{TA,k}}$};

\draw (1,0.01) -- (1,-0.01);
\draw (1.5,0.01) -- (1.5,-0.01);
\draw (2,0.01) -- (2,-0.01);

\draw (1.01,0.333) -- (0.99,0.333);
\end{tikzpicture}
\vspace{-2mm}
\caption{Relative error profile within interval $[V_k, V_{k+1})$. The solid line shows the error profile computed by integration ($\text{CRE}_{\text{IM,k}}$); the dashed line shows the triangle approximation ($\text{CRE}_{\text{TA,k}}$). The shaded area represents the cumulative relative error ($\text{CRE}_{\text{k}}$).}
\label{fig:mountain_diagram}
\end{figure}

Figure~\ref{fig:mountain_diagram} illustrates the relative error profile within a single interval $[V_k, V_{k+1})$. For any true value $V$ in the interval $[V_k, V_{k+1})$, it rounds to the nearest representable WINT value. The decision boundary is the midpoint $V_{\text{half}} = (V_k + V_{k+1})/2$. Values below $V_{\text{half}}$ round to $V_k$ and values above round to $V_{k+1}$. The relative error of representing any point $V \in [V_k, V_{k+1})$
  is given by Equation~\ref{eq:relative_error}, which yields:
\begin{equation}
\small
\text{RE}(V) = \begin{cases}
\dfrac{V - V_k}{V} & \text{if } V_k \leq V < V_{\text{half}} \\[10pt]
\dfrac{V_{k+1} - V}{V} & \text{if } V_{\text{half}} \leq V \leq V_{k+1}
\end{cases}
\label{eq:re_piecewise}
\end{equation}

As shown in Figure~\ref{fig:mountain_diagram}, the RE profile within each interval forms a characteristic shape: the left half is concave (bulging above a straight line) and the right half is convex (sagging below). This is because the RE expression is inversely proportional to $V$. Both halves meet at the midpoint where relative error reaches its maximum $RE_{max}$, and both decrease to zero at the interval endpoints.

The cumulative relative error for one interval is obtained by integrating over both halves of the interval.
The cumulative relative error for interval $k$ is $\text{CRE}_k = \int_{V_k}^{V_{\text{half}}} \frac{V - V_k}{V} \, dV + \int_{V_{\text{half}}}^{V_{k+1}} \frac{V_{k+1} - V}{V} \, dV$, which evaluates to:
\begin{equation}
\text{CRE}_k = V_{k+1} \ln\!\bigl(\tfrac{V_{k+1}}{V_{\text{half}}}\bigr) - V_k \ln\!\bigl(\tfrac{V_{\text{half}}}{V_k}\bigr)
\label{eq:cre_result}
\end{equation}

The MRE is computed by summing the $\text{CRE}_k$ over all $N-1$ intervals (where $N$ is the total number of unique representable values) and dividing by the maximum representable value, $V_{\text{range}}$:
\begin{equation}
\small
\text{MRE} = \frac{1}{V_{\text{range}}} \sum_{k=0}^{N-1} \text{CRE}_k
\label{eq:mre_integral}
\end{equation}

While the IM produces accurate results, it requires evaluating the natural logarithm (in Equation~\ref{eq:cre_result}) for every interval. Since the number of intervals grows exponentially with $m$, the IM-based MRE is prohibitively slow for large mantissa sizes. Also, each interval requires logarithmic evaluations which demand high precision arithmetic to maintain accuracy.

\subsection{Triangle Approximation (TA)}
\label{subsec:triangle_method}

The IM requires evaluating $\text{CRE}_k$ (Equation~\ref{eq:cre_result}) for every interval, which involves logarithmic computations at high precision. The Triangle Approximation (TA) method addresses this computational bottleneck by approximating the curved error profile (solid line of Figure~\ref{fig:mountain_diagram}) with a simpler geometric shape (dashed line of Figure~\ref{fig:mountain_diagram}).

As shown in Figure~\ref{fig:mountain_diagram}, the relative error within an interval $[V_k, V_{k+1})$ rises from zero at the endpoints to a maximum at the midpoint, forming a roughly triangular shape. Since the true error profile (the solid line in Figure~\ref{fig:mountain_diagram}) closely resembles a triangle (dashed line in Figure~\ref{fig:mountain_diagram}), we approximate the cumulative relative error as the area of this triangle ($\text{CRE}_{\text{TA,k}}$):
\begin{equation}
\small
\text{CRE}_{\text{TA,k}} = \frac{1}{2} (V_{k+1} - V_k) \times \text{RE}_{\max}
\label{eq:triangle_area}
\end{equation}

To evaluate this for WINT, we denote a specific mantissa value as $m_i$ (where $0 \leq m_i \leq 2^m - 1$) and a specific exponent value as $e_j$ (where $0 \leq e_j \leq 2^e - 1$). A WINT value is then $V = m_i \times 2^{e_j}$. Consecutive mantissa values at the same exponent level differ by $2^{e_j}$, so the interval width (base of the triangle) equals $2^{e_j}$. The height $\text{RE}_{\max}$ occurs at the midpoint $V_{\text{half}} = (m_i + 0.5) \times 2^{e_j}$, which yields $\text{RE}_{\max} = \tfrac{V_{\text{half}} - V_k}{V_{\text{half}}} = \tfrac{0.5 \times 2^{e_j}}{(m_i + 0.5) \times 2^{e_j}} = \tfrac{0.5}{m_i + 0.5}$. The exponent cancels, leaving $\text{RE}_{\max}$ dependent only on the mantissa. Substituting into Equation~\ref{eq:triangle_area}, we get:
\vspace{-4mm}
\begin{equation}
\small
\text{CRE}_{\text{TA,k}} \approx \frac{1}{2} \cdot 2^{e_j} \cdot \frac{0.5}{m_i + 0.5} = \frac{2^{e_j-2}}{m_i + 0.5}
\label{eq:cre_triangle}
\end{equation}

The total MRE sums this contribution over all unique intervals. However, not every $(m_i, e_j)$ combination produces a unique value, as noted in Section~\ref{subsec:wint_definition}. Table~\ref{tab:wint_values} shows all representable values for a $\{m=4, e=2\}$ WINT configuration, revealing which combinations are unique and which are duplicates.

\begin{table*}[!t]
\centering
\small
\caption{Representable values for WINT $\{m=4, e=2\}$.}
\vspace{-1mm}
\label{tab:wint_values}

\renewcommand{\arraystretch}{0.92} 
\setlength{\tabcolsep}{3.5pt}      

\begin{tabular}{c|*{16}{c}}
\toprule
& \multicolumn{16}{c}{Mantissa $m_i$ (0--15)} \\
\textbf{Exponent $e_j$ (0--3)} & 0 & 1 & 2 & 3 & 4 & 5 & 6 & 7 & 8 & 9 & 10 & 11 & 12 & 13 & 14 & 15 \\
\midrule
0 & \textbf{0} & \textbf{1} & \textbf{2} & \textbf{3} & \textbf{4} & \textbf{5} & \textbf{6} & \textbf{7} & \textit{8} & \textit{9} & \textit{10} & \textit{11} & \textit{12} & \textit{13} & \textit{14} & \textit{15} \\
1 & (0) & (2) & (4) & (6) & (8) & (10) & (12) & (14) & \textit{16} & \textit{18} & \textit{20} & \textit{22} & \textit{24} & \textit{26} & \textit{28} & \textit{30} \\
2 & (0) & (4) & (8) & (12) & (16) & (20) & (24) & (28) & \textit{32} & \textit{36} & \textit{40} & \textit{44} & \textit{48} & \textit{52} & \textit{56} & \textit{60} \\
3 & (0) & (8) & (16) & (24) & (32) & (40) & (48) & (56) & \textit{64} & \textit{72} & \textit{80} & \textit{88} & \textit{96} & \textit{104} & \textit{112} & \textit{120} \\
\bottomrule
\end{tabular}
\vspace{-1mm}
\end{table*}

The table reveals two distinct regions. The \textbf{bold region} contains mantissa values $m_i = 0$ to $2^{m-1}-1$ at $e_j = 0$, which includes values that are unique. The \textit{italicized region} contains mantissa values $m_i = 2^{m-1}$ to $2^m - 1$, which generate new unique values for every exponent. From Equation~\ref{eq:cre_triangle}, iterating over all mantissa-exponent pairs and summing over unique intervals gives:
\vspace{-2mm}
\begin{equation}
\small
\text{MRE} = \frac{1}{V_{\text{range}}} \sum_{j=0}^{2^e-1} \sum_{i=m_{e_j}}^{2^m-1} \frac{2^{e_j-2}}{m_i + 0.5}
\label{eq:mre_double_sum}
\end{equation}
where $m_{e_j}$ is the starting mantissa index for exponent $e_j$. From Table~\ref{tab:wint_values}, we can see that $m_{e_j} = 1$ for $e_j = 0$ (bold region), and $m_{e_j} = 2^{m-1}$ for $e_j \geq 1$ (italicized region).

The exponent factor $2^{e_j-2}$ can be factored out and evaluated as a geometric series. However, the mantissa summation $\sum \frac{1}{m_i + 0.5}$ has no closed-form solution. Splitting based on unique $(m_i, e_j)$ pairs and evaluating the geometric series yields:
\vspace{-3mm}
\begin{equation}
\small
\begin{aligned}
\mathrm{MRE} = \frac{1}{V_{\text{range}}} \Bigg[
&\,0.25 \sum_{m_i=1}^{2^{m-1}-1} \frac{1}{m_i + 0.5} \\
&+ \left(2^{2^e-2} - 0.25\right)
  \sum_{m_i=2^{m-1}}^{2^m-1}
  \frac{1}{m_i + 0.5}
\Bigg]
\end{aligned}
\label{eq:mre_triangle_final}
\end{equation}

The bold region corresponds to the first term, while the italicized region corresponds to the second term.

The exponent contributions are now closed-form, but the mantissa summations still require iterating over $2^{m-1}$ terms each. For large $m$, this remains computationally expensive, motivating the approximation methods that follow.

From the structure of Table~\ref{tab:wint_values}, the number of unique representable values for a given WINT configuration $\{m, e\}$ can be determined. The bold region contributes $2^{m-1}$ values (for $e_j = 0$ only), while the italicized region contributes $2^{m-1}$ new values for each of the $2^e$ exponent levels, giving $2^{m-1} \times (2^e + 1)$ unique values.

Both the IM and TA methods require iterating over all mantissa values, which means summing $2^m - 1$ terms. For configurations such as $32$-bit WINT with $m = 24$, this means evaluating over $16$ million terms. This becomes intractable especially since high precision arithmetic is needed to maintain accuracy. This motivates the development of approximation methods that eliminate the explicit summation over mantissa values.

\subsection{Taylor Series Approximation (TSA)}
\label{subsec:taylor_approx}

The mantissa summations in Equation~\ref{eq:mre_triangle_final} involve sums of the form $\sum \frac{1}{m_i + 0.5}$, which must be evaluated term by term in the TA method. The Taylor series approach eliminates this explicit iteration by approximating the reciprocal function with a polynomial expansion, converting the summation into closed-form power sums.

The Taylor series expansion can approximate the reciprocal function, enabling closed-form evaluation of the mantissa summations. For $\frac{1}{x}$ expanded around point $a$:

\begin{equation}
\small
\frac{1}{x} = \sum_{n=0}^{\infty} \frac{(-1)^n (x-a)^n}{a^{n+1}}, \quad |x-a| < a
\label{eq:taylor_1x}
\end{equation}

To approximate $\sum_{i=1}^{N} \frac{1}{m_i+0.5}$, we substitute $x = m_i + 0.5$ and apply the expansion. Using the first six terms and simplifying:
\vspace{-2mm}
\begin{equation}
\small
\frac{1}{x} \approx \frac{6}{a} - \frac{15x}{a^2} + \frac{20x^2}{a^3} - \frac{15x^3}{a^4} + \frac{6x^4}{a^5} - \frac{x^5}{a^6}
\label{eq:taylor_expanded}
\end{equation}

Summing over all values converts each term to a power sum:
\vspace{-2mm}
\begin{equation}
\small
\sum_{i=1}^{N} \frac{1}{m_i+0.5} \approx \frac{6}{a}\sum_{i=1}^{N}1 - \frac{15}{a^2}\sum_{i=1}^{N}(m_i+0.5) + \cdots
\label{eq:taylor_power_sums}
\end{equation}

These power sums $\sum_{i=1}^{N} i^k$ have well-known closed-form expressions for higher powers up to $k=5$.

However, a single expansion point $a$ cannot accurately approximate $\frac{1}{x}$ across the entire mantissa range $[1, 2^m)$, as values far from $a$ violate the convergence criterion $|x-a| < a$. To address this, we partition the range into segments where values approximately double: $[1, 2), [2, 4), [4, 8), \ldots, [2^{m-1}, 2^m)$.

For each segment $[2^k, 2^{k+1})$, we choose expansion point $a = 1.5 \times 2^k$, the segment midpoint. This ensures all values within the segment satisfy the convergence criterion while minimizing approximation error. Ten terms were used in calculations, but using five or more does not significantly change accuracy.

This segmentation reduces the computational complexity significantly, with each segment requiring only constant-time evaluation of the closed-form power sums. While the TSA method achieves O(1) time complexity, it requires careful selection of expansion points and generally exhibits slightly larger approximation errors than the harmonic approximation method (described next).

\subsection{Harmonic Approximation (HA)}
\label{subsec:harmonic_approx}

The harmonic approximation takes a different approach to eliminating the explicit mantissa summation. Rather than approximating the reciprocal function term by term as in the Taylor method, it recognizes that the mantissa summations in Equation~\ref{eq:mre_triangle_final} can be expressed in terms of harmonic numbers, which in turn have efficient asymptotic approximations. The harmonic series is defined as the sum of reciprocals of positive integers:
\vspace{-2mm}
\begin{equation}
\small
\sum_{i=1}^{\infty} \frac{1}{i} = 1 + \frac{1}{2} + \frac{1}{3} + \frac{1}{4} + \cdots
\label{eq:harmonic_series}
\end{equation}
\vspace{-3mm}

The $n$-th harmonic number $H_n$ is the partial sum:
\begin{equation}
\small
H_n = \sum_{i=1}^{n} \frac{1}{i}
\label{eq:harmonic_number}
\end{equation}
\vspace{-1mm}

For large $n$, $H_n$ can be approximated using the Euler-Maclaurin expansion~\cite{euler_maclaurin_citation}:
\vspace{-2mm}
\begin{equation}
\small
H_n \approx \ln(n) + \gamma + \frac{1}{2n} - \sum_{k=1}^{\infty} \frac{B_{2k}}{2k \cdot n^{2k}}
\label{eq:euler_maclaurin}
\end{equation}
where $\gamma \approx 0.5772156649$ is the Euler-Mascheroni constant and $B_{2k}$ are the Bernoulli numbers.

\subsubsection{Converting to Harmonic Numbers}

The key insight is recognizing that the mantissa summations in Equation~\ref{eq:mre_triangle_final} can be transformed into sums of odd reciprocals:
\begin{equation}
\small
\sum_{i=1}^{N} \frac{1}{m_i + 0.5} = 2 \sum_{i=1}^{N} \frac{1}{2m_i+1}
\label{eq:to_odd_reciprocals}
\end{equation}

The sum of odd reciprocals relates to harmonic numbers via the identity:
\vspace{-2mm}
\begin{equation}
\small
H_{2N+1} = \sum_{i=0}^{N} \frac{1}{2i+1} + \frac{1}{2}H_N
\label{eq:harmonic_odd_identity}
\end{equation}

Rearranging and adjusting for index starting at $i=1$:
\vspace{-1mm}
\begin{equation}
\small
\sum_{i=1}^{N} \frac{1}{2i+1} = H_{2N+1} - \frac{1}{2}H_N - 1
\label{eq:odd_sum_result}
\end{equation}

Combining with Equation~\ref{eq:to_odd_reciprocals}, we obtain the key result:
\vspace{-1mm}
\begin{equation}
\small
\sum_{i=1}^{N} \frac{1}{i+0.5} = 2H_{2N+1} - H_N - 2
\label{eq:sum_to_harmonic}
\end{equation}

This identity allows us to express both mantissa summations in Equation~\ref{eq:mre_triangle_final} in terms of harmonic numbers. Since $H_n$ can be approximated via the Euler-Maclaurin expansion using only a fixed number of arithmetic operations regardless of $n$, the harmonic approximation also achieves O(1) time complexity.

\subsubsection{Selecting Number of Terms}

The Euler-Maclaurin expansion is an \textit{asymptotic series}~\cite{euler_maclaurin_citation}---it approaches the true value up to an optimal truncation point, then diverges. This occurs because Bernoulli numbers grow factorially, eventually overwhelming the polynomial denominator.

For the series to improve with additional terms, the denominator growth ($n^{2k}$) must exceed numerator growth ($|B_{2k}|$). This requires sufficiently large $n = 2^m$. Section~\ref{subsec:harmonic_validation} empirically determines the optimal number of terms for our problem.

\section{Results}
\label{sec:results}

This section presents experimental validation of the approximation methods and comprehensive characterization of WINT across multiple bit widths, with a comparison against the standard integer. Section~\ref{subsec:experimental_setup} describes the experimental setup. Section~\ref{subsec:harmonic_validation} validates the harmonic approximation against exact calculations. Section~\ref{subsec:method_comparison} compares the MRE computation of all four computation methods. Section~\ref{subsec:wint_tradeoff} explores WINT's precision-range trade-offs.
\vspace{-2mm}
\subsection{Experimental Setup}
\label{subsec:experimental_setup}

This subsection describes the hardware and software environment used for all experiments. Experiments were conducted on a high-performance computing cluster. The compute node used is equipped with dual AMD EPYC 7F72 processors (24 cores per socket, 3.2\,GHz base clock) and 1\,TB of RAM, running Red Hat Enterprise Linux 8.10. All calculations were implemented in Python 3.13.5 using the mpmath library~\cite{mpmath_citation} with 100 decimal places of precision.

\vspace{-1mm}

\subsection{Harmonic Approximation Validation}
\label{subsec:harmonic_validation}

To measure the correctness of the harmonic approximation under various mantissa sizes, we validated its accuracy by examining the individual mantissa summations. Recall from Equation~\ref{eq:mre_triangle_final} that the MRE computation requires evaluating sums of the form $\sum_{i=1}^{N} \frac{1}{m_i + 0.5}$. The TA method computes these sums explicitly, while the HA method uses the identity in Equation~\ref{eq:sum_to_harmonic} combined with the Euler-Maclaurin expansion to approximate them. We measured the percentage difference between the result of the HA method and the TA method for $\sum_{i=1}^{2^{m-1}-1} \frac{1}{m_i + 0.5}$ to determine the optimal number of Bernoulli correction terms.

Table~\ref{tab:term_selection} shows how the approximation error varies with the number of Bernoulli correction terms for different mantissa sizes. For $m \geq 5$, the approximation converges rapidly and remains stable above 7 terms. However, for $m = 4$, the series exhibits the asymptotic divergence discussed in Section~\ref{subsec:harmonic_approx}: accuracy improves initially but degrades for 20 or more terms as Bernoulli number growth overwhelms the denominator of Equation~\ref{eq:euler_maclaurin}.

\begin{table}[t]
\centering
\caption{Approximation error (\%) for the mantissa summation across different term counts.}
\vspace{-2mm}
\label{tab:term_selection}
\resizebox{\columnwidth}{!}{
\begin{tabular}{c|cccc}
\toprule
\# Terms & $m=4$ & $m=5$ & $m=8$ & $m=16$ \\
\midrule
1  & $1.5 \times 10^{-4}$ & $5.3 \times 10^{-6}$ & $5.8 \times 10^{-10}$ & $4.8 \times 10^{-17}$ \\
2  & $1.6 \times 10^{-6}$ & $1.2 \times 10^{-8}$ & $1.9 \times 10^{-14}$ & $4.8 \times 10^{-17}$ \\
3  & $3.4 \times 10^{-8}$ & $5.9 \times 10^{-11}$ & $1.0 \times 10^{-16}$ & $4.8 \times 10^{-17}$ \\
4  & $1.2 \times 10^{-9}$ & $4.7 \times 10^{-13}$ & $1.0 \times 10^{-16}$ & $4.8 \times 10^{-17}$ \\
5  & $6.9 \times 10^{-11}$ & $6.0 \times 10^{-15}$ & $1.0 \times 10^{-16}$ & $4.8 \times 10^{-17}$ \\
6  & $5.4 \times 10^{-12}$ & $7.9 \times 10^{-17}$ & $1.0 \times 10^{-16}$ & $4.8 \times 10^{-17}$ \\
7  & $5.7 \times 10^{-13}$ & $1.8 \times 10^{-16}$ & $1.0 \times 10^{-16}$ & $4.8 \times 10^{-17}$ \\
8  & $7.8 \times 10^{-14}$ & $1.8 \times 10^{-16}$ & $1.0 \times 10^{-16}$ & $4.8 \times 10^{-17}$ \\
9  & $1.4 \times 10^{-14}$ & $1.8 \times 10^{-16}$ & $1.0 \times 10^{-16}$ & $4.8 \times 10^{-17}$ \\
10 & $2.6 \times 10^{-15}$ & $1.8 \times 10^{-16}$ & $1.0 \times 10^{-16}$ & $4.8 \times 10^{-17}$ \\
20 & $2.4 \times 10^{-16}$ & $1.8 \times 10^{-16}$ & $1.0 \times 10^{-16}$ & $4.8 \times 10^{-17}$ \\
40 & $2.3 \times 10^{-13}$ & $1.8 \times 10^{-16}$ & $1.0 \times 10^{-16}$ & $4.8 \times 10^{-17}$ \\
60 & $3.6 \times 10^{-7}$ & $1.8 \times 10^{-16}$ & $1.0 \times 10^{-16}$ & $4.8 \times 10^{-17}$ \\
80 & $3.6 \times 10^{-7}$ & $1.8 \times 10^{-16}$ & $1.0 \times 10^{-16}$ & $4.8 \times 10^{-17}$ \\
\bottomrule
\end{tabular}
}
\vspace{-2mm}
\end{table}

Based on this analysis, we selected 10 Bernoulli correction terms for all subsequent experiments. This provides stable convergence across all $m \geq 5$ configurations. For $m = 4$, the harmonic approximation achieves its best accuracy around 10--20 terms before diverging. We note that we can just use integral calculation to get MRE results for configurations with low m values since the small mantissa range makes direct summation fast.
\vspace{-2mm}
\subsection{Method Comparison}
\label{subsec:method_comparison}

This subsection compares all four MRE calculation methods---IM, TA, TSA, and HA---across the full range of configurations. For each bit width, we present WINT configurations (where $e \geq 1$) separately from standard integer baselines ($e = 0$), since configurations with zero exponent bits are functionally equivalent to standard unsigned integers. For brevity, each table presents only configurations where the MRE values of WINT are comparable to the integer baseline, focusing on small exponent allocations ($e \leq 5$) that are most relevant for practical applications.

Tables~\ref{tab:results_8bit}--\ref{tab:results_32bit} present results for 8-bit through 32-bit configurations. Each table shows the MRE values, computation times, and maximum representable range for all four methods. The error percentage columns show the deviation from the IM calculation (the exact method). Several key findings emerge from these results:

\textbf{Computational speedup.} The HA method maintains O(1) time complexity of approximately 0.0005 seconds up to 32 bits, while the IM and TA methods' runtimes increase exponentially as $m$ increases. For the 32-bit configuration with \{$m=31$, $e=1$\}, the IM calculation required 143,848 seconds (approximately 40 hours), while the HA method completed in 0.0006 seconds---a speedup of approximately $2.4 \times 10^{8}$. The TSA method also achieves O(1) time complexity but is slower than HA by a factor of approximately 100.

\textbf{Method accuracy comparison.} All three approximation methods achieve excellent accuracy. The TA and HA methods produce identical error percentages because they use the same underlying approximation---the HA method simply provides a closed-form evaluation of the TA method's summation. The TSA method shows slightly larger errors for mid-range mantissa values, while matching the other methods for configurations where $m$ approaches the total bit width.

We also computed MRE for 48-bit, 64-bit, and 128-bit configurations using all four methods. These results are omitted from the tables because computing the integer baseline ($e = 0$) via the integral method was infeasible at these bit widths, as it would require weeks or longer. The approximation methods continued to perform consistently at these larger bit widths: HA and TSA maintained O(1) time complexity, and the accuracy patterns observed for 8--32 bit configurations held across all tested configurations up to 128 bits.

\vspace{-2mm}

\subsection{WINT Tradeoff Exploration}
\label{subsec:wint_tradeoff}
We now use the results from Tables~\ref{tab:results_8bit}-~\ref{tab:results_32bit} to characterize WINT's trade-offs. The Max Range column in each table shows the maximum representable value (Equation~\ref{eq:wint_max}) for each configuration, enabling a direct comparison of both error and range against the integer baseline.

The tables reveal several important patterns. Across all bit widths from 12 to 32, allocating 2 exponent bits doubles
the representable range while simultaneously reducing MRE by 12--33\% compared to the integer baseline. For example,
12-bit WINT with $\{m = 10, e = 2\}$ achieves a range of $8.18 \times 10^{3}$ (versus $4.10 \times 10^{3}$ for the
baseline) with MRE of $5.92 \times 10^{-4}$ (12\% lower than the baseline's $6.76 \times 10^{-4}$), while 32-bit WINT
with $\{m = 30, e = 2\}$ achieves a range of $8.59 \times 10^{9}$ (versus $4.29 \times 10^{9}$) with MRE of
$9.68 \times 10^{-10}$ (33\% lower than the baseline's $1.45 \times 10^{-9}$).

Allocating 3 exponent bits extends the range by $16\times$ while reducing MRE by 15--50\% for bit widths of 16 and
above. For instance, 32-bit WINT with $\{m = 29, e = 3\}$ achieves a range of $6.87 \times 10^{10}$ ($16\times$ the
baseline) with MRE of $7.24 \times 10^{-10}$ (50\% lower). With 4 exponent bits, the range increases by over $2000\times$,
though the MRE benefit diminishes---at 32 bits, $\{m = 28, e = 4\}$ provides a range of $8.80 \times 10^{12}$ with MRE
of $1.29 \times 10^{-9}$ (11\% lower than the baseline). Allocating only a single exponent bit ($e = 1$) provides
essentially the same maximum range as the integer baseline while slightly increasing MRE, because the mantissa loses
one bit of resolution without a meaningful gain in range.

Users seeking a balance of range extension and accuracy improvement should therefore allocate 2--3 exponent bits for
their target application. Importantly, WINT is able to offer configurations with a larger range \emph{and} lower MRE than the standard integer representation.


\begin{table*}[!t]
 
\centering
\captionof{table}{Method comparison for 8-bit configurations. Times are in seconds.}
\vspace{-2mm}
\label{tab:results_8bit}
\renewcommand{\arraystretch}{0.99}
\resizebox{0.94\textwidth}{!}{
\begin{tabular}{c|cc|r|cc|ccc|ccc|ccc}
\toprule
& & & & \multicolumn{2}{c|}{IM Method} & \multicolumn{3}{c|}{TA Method} & \multicolumn{3}{c|}{TSA Method} & \multicolumn{3}{c}{HA Method} \\
& $m$ & $e$ & Max Range & MRE & Time & MRE & Time & Err\% & MRE & Time & Err\% & MRE & Time & Err\% \\
\midrule
\multirow{4}{*}{\textit{WINT}}
& 4 & 4 & $4.92 \times 10^{5}$ & $2.20 \times 10^{-2}$ & 0.0013 & $2.20 \times 10^{-2}$ & 0.0002 & 0.036 & $2.20 \times 10^{-2}$ & 0.0130 & 0.036 & $2.20 \times 10^{-2}$ & 0.0005 & 0.036 \\
& 5 & 3 & $3.97 \times 10^{3}$ & $1.12 \times 10^{-2}$ & 0.0023 & $1.12 \times 10^{-2}$ & 0.0003 & 0.019 & $1.12 \times 10^{-2}$ & 0.0162 & 0.019 & $1.12 \times 10^{-2}$ & 0.0005 & 0.019 \\
& 6 & 2 & $5.04 \times 10^{2}$ & $8.18 \times 10^{-3}$ & 0.0043 & $8.17 \times 10^{-3}$ & 0.0005 & 0.110 & $8.17 \times 10^{-3}$ & 0.0195 & 0.110 & $8.17 \times 10^{-3}$ & 0.0005 & 0.110 \\
& 7 & 1 & $2.54 \times 10^{2}$ & $8.84 \times 10^{-3}$ & 0.0086 & $8.82 \times 10^{-3}$ & 0.0009 & 0.200 & $8.82 \times 10^{-3}$ & 0.0228 & 0.200 & $8.82 \times 10^{-3}$ & 0.0005 & 0.200 \\
\midrule
\textit{Int}
& 8 & 0 & $2.55 \times 10^{2}$ & $8.13 \times 10^{-3}$ & 0.0173 & $8.12 \times 10^{-3}$ & 0.0019 & 0.216 & $8.12 \times 10^{-3}$ & 0.0260 & 0.216 & $8.12 \times 10^{-3}$ & 0.0005 & 0.216 \\
\bottomrule
\end{tabular}
}
\vspace{-1mm}
\centering
\captionof{table}{Method comparison for 12-bit configurations. Times are in seconds.}
\label{tab:results_12bit}
\renewcommand{\arraystretch}{0.99}
\resizebox{0.94\textwidth}{!}{
\begin{tabular}{c|cc|r|cc|ccc|ccc|ccc}
\toprule
& & & & \multicolumn{2}{c|}{IM Method} & \multicolumn{3}{c|}{TA Method} & \multicolumn{3}{c|}{TSA Method} & \multicolumn{3}{c}{HA Method} \\
& $m$ & $e$ & Max Range & MRE & Time & MRE & Time & Err\% & MRE & Time & Err\% & MRE & Time & Err\% \\
\midrule
\multirow{4}{*}{\textit{WINT}}
& 8 & 4 & $8.36 \times 10^{6}$ & $1.36 \times 10^{-3}$ & 0.0173 & $1.36 \times 10^{-3}$ & 0.0019 & $1.8 \times 10^{-4}$ & $1.36 \times 10^{-3}$ & 0.0259 & $3.3 \times 10^{-4}$ & $1.36 \times 10^{-3}$ & 0.0005 & $1.8 \times 10^{-4}$ \\
& 9 & 3 & $6.54 \times 10^{4}$ & $7.06 \times 10^{-4}$ & 0.0349 & $7.06 \times 10^{-4}$ & 0.0038 & 0.010 & $7.06 \times 10^{-4}$ & 0.0292 & 0.010 & $7.06 \times 10^{-4}$ & 0.0005 & 0.010 \\
& 10 & 2 & $8.18 \times 10^{3}$ & $5.92 \times 10^{-4}$ & 0.0706 & $5.92 \times 10^{-4}$ & 0.0076 & 0.092 & $5.92 \times 10^{-4}$ & 0.0324 & 0.093 & $5.92 \times 10^{-4}$ & 0.0005 & 0.092 \\
& 11 & 1 & $4.09 \times 10^{3}$ & $7.18 \times 10^{-4}$ & 0.1414 & $7.17 \times 10^{-4}$ & 0.0152 & 0.152 & $7.17 \times 10^{-4}$ & 0.0355 & 0.152 & $7.17 \times 10^{-4}$ & 0.0005 & 0.152 \\
\midrule
\textit{Int}
& 12 & 0 & $4.10 \times 10^{3}$ & $6.76 \times 10^{-4}$ & 0.2816 & $6.75 \times 10^{-4}$ & 0.0305 & 0.162 & $6.75 \times 10^{-4}$ & 0.0387 & 0.162 & $6.75 \times 10^{-4}$ & 0.0005 & 0.162 \\
\bottomrule
\end{tabular}
}

\vspace{-1mm}
\centering
\captionof{table}{Method comparison for 16-bit configurations. Times are in seconds.}

\label{tab:results_16bit}
\renewcommand{\arraystretch}{0.99}
\resizebox{0.94\textwidth}{!}{
\begin{tabular}{c|cc|r|cc|ccc|ccc|ccc}
\toprule
& & & & \multicolumn{2}{c|}{IM Method} & \multicolumn{3}{c|}{TA Method} & \multicolumn{3}{c|}{TSA Method} & \multicolumn{3}{c}{HA Method} \\
& $m$ & $e$ & Max Range & MRE & Time & MRE & Time & Err\% & MRE & Time & Err\% & MRE & Time & Err\% \\
\midrule
\multirow{5}{*}{\textit{WINT}}
& 11 & 5 & $4.40 \times 10^{12}$ & $1.69 \times 10^{-4}$ & 0.1415 & $1.69 \times 10^{-4}$ & 0.0153 & $2.2 \times 10^{-6}$ & $1.69 \times 10^{-4}$ & 0.0356 & $1.6 \times 10^{-4}$ & $1.69 \times 10^{-4}$ & 0.0005 & $2.2 \times 10^{-6}$ \\
& 12 & 4 & $1.34 \times 10^{8}$ & $8.46 \times 10^{-5}$ & 0.2824 & $8.46 \times 10^{-5}$ & 0.0304 & $4.0 \times 10^{-5}$ & $8.46 \times 10^{-5}$ & 0.0394 & $2.0 \times 10^{-4}$ & $8.46 \times 10^{-5}$ & 0.0006 & $4.0 \times 10^{-5}$ \\
& 13 & 3 & $1.05 \times 10^{6}$ & $4.48 \times 10^{-5}$ & 0.5646 & $4.48 \times 10^{-5}$ & 0.0606 & 0.010 & $4.48 \times 10^{-5}$ & 0.0423 & 0.010 & $4.48 \times 10^{-5}$ & 0.0005 & 0.010 \\
& 14 & 2 & $1.31 \times 10^{5}$ & $4.23 \times 10^{-5}$ & 1.12 & $4.22 \times 10^{-5}$ & 0.1219 & 0.081 & $4.22 \times 10^{-5}$ & 0.0455 & 0.081 & $4.22 \times 10^{-5}$ & 0.0005 & 0.081 \\
& 15 & 1 & $6.55 \times 10^{4}$ & $5.55 \times 10^{-5}$ & 2.26 & $5.54 \times 10^{-5}$ & 0.2443 & 0.123 & $5.54 \times 10^{-5}$ & 0.0486 & 0.123 & $5.54 \times 10^{-5}$ & 0.0005 & 0.123 \\
\midrule
\textit{Int}
& 16 & 0 & $6.55 \times 10^{4}$ & $5.28 \times 10^{-5}$ & 4.54 & $5.27 \times 10^{-5}$ & 0.4898 & 0.129 & $5.27 \times 10^{-5}$ & 0.0523 & 0.129 & $5.27 \times 10^{-5}$ & 0.0006 & 0.129 \\
\bottomrule
\end{tabular}
}

\vspace{-1mm}
\centering
\captionof{table}{Method comparison for 24-bit configurations. Times are in seconds.}

\label{tab:results_24bit}
\renewcommand{\arraystretch}{0.99}
\resizebox{0.94\textwidth}{!}{
\begin{tabular}{c|cc|r|cc|ccc|ccc|ccc}
\toprule
& & & & \multicolumn{2}{c|}{IM Method} & \multicolumn{3}{c|}{TA Method} & \multicolumn{3}{c|}{TSA Method} & \multicolumn{3}{c}{HA Method} \\
& $m$ & $e$ & Max Range & MRE & Time & MRE & Time & Err\% & MRE & Time & Err\% & MRE & Time & Err\% \\
\midrule
\multirow{5}{*}{\textit{WINT}}
& 19 & 5 & $1.13 \times 10^{15}$ & $6.61 \times 10^{-7}$ & 35.9 & $6.61 \times 10^{-7}$ & 3.89 & $6.3 \times 10^{-10}$ & $6.61 \times 10^{-7}$ & 0.0618 & $1.6 \times 10^{-4}$ & $6.61 \times 10^{-7}$ & 0.0006 & $6.3 \times 10^{-10}$ \\
& 20 & 4 & $3.44 \times 10^{10}$ & $3.31 \times 10^{-7}$ & 71.7 & $3.31 \times 10^{-7}$ & 7.83 & $3.9 \times 10^{-5}$ & $3.31 \times 10^{-7}$ & 0.0652 & $2.0 \times 10^{-4}$ & $3.31 \times 10^{-7}$ & 0.0006 & $3.9 \times 10^{-5}$ \\
& 21 & 3 & $2.68 \times 10^{8}$ & $1.80 \times 10^{-7}$ & 142.9 & $1.80 \times 10^{-7}$ & 15.8 & 0.009 & $1.80 \times 10^{-7}$ & 0.0682 & 0.009 & $1.80 \times 10^{-7}$ & 0.0006 & 0.009 \\
& 22 & 2 & $3.36 \times 10^{7}$ & $2.06 \times 10^{-7}$ & 285.5 & $2.06 \times 10^{-7}$ & 31.4 & 0.065 & $2.06 \times 10^{-7}$ & 0.0726 & 0.065 & $2.06 \times 10^{-7}$ & 0.0006 & 0.065 \\
& 23 & 1 & $1.68 \times 10^{7}$ & $2.99 \times 10^{-7}$ & 569.0 & $2.99 \times 10^{-7}$ & 61.6 & 0.089 & $2.99 \times 10^{-7}$ & 0.0747 & 0.089 & $2.99 \times 10^{-7}$ & 0.0006 & 0.089 \\
\midrule
\textit{Int}
& 24 & 0 & $1.68 \times 10^{7}$ & $2.89 \times 10^{-7}$ & 1139 & $2.89 \times 10^{-7}$ & 124.2 & 0.092 & $2.89 \times 10^{-7}$ & 0.0786 & 0.092 & $2.89 \times 10^{-7}$ & 0.0006 & 0.092 \\
\bottomrule
\end{tabular}
}

\vspace{-1mm}
\centering
\captionof{table}{Method comparison for 32-bit configurations. Times are in seconds.}

\label{tab:results_32bit}
\renewcommand{\arraystretch}{0.99}
\resizebox{0.94\textwidth}{!}{
\begin{tabular}{c|cc|r|cc|ccc|ccc|ccc}
\toprule
& & & & \multicolumn{2}{c|}{IM Method} & \multicolumn{3}{c|}{TA Method} & \multicolumn{3}{c|}{TSA Method} & \multicolumn{3}{c}{HA Method} \\
& $m$ & $e$ & Max Range & MRE & Time & MRE & Time & Err\% & MRE & Time & Err\% & MRE & Time & Err\% \\
\midrule
\multirow{5}{*}{\textit{WINT$^{*}$}}
& 27 & 5 & $2.88 \times 10^{17}$ & $2.58 \times 10^{-9}$ & 9121 & $2.58 \times 10^{-9}$ & 982 & $6.0 \times 10^{-10}$ & $2.58 \times 10^{-9}$ & 0.0867 & $1.6 \times 10^{-4}$ & $2.58 \times 10^{-9}$ & 0.0006 & $6.0 \times 10^{-10}$ \\
& 28 & 4 & $8.80 \times 10^{12}$ & $1.29 \times 10^{-9}$ & 18190 & $1.29 \times 10^{-9}$ & 1963 & $3.9 \times 10^{-5}$ & $1.29 \times 10^{-9}$ & 0.0896 & $2.0 \times 10^{-4}$ & $1.29 \times 10^{-9}$ & 0.0006 & $3.9 \times 10^{-5}$ \\
& 29 & 3 & $6.87 \times 10^{10}$ & $7.24 \times 10^{-10}$ & 35935 & $7.24 \times 10^{-10}$ & 3922 & 0.009 & $7.24 \times 10^{-10}$ & 0.0930 & 0.009 & $7.24 \times 10^{-10}$ & 0.0006 & 0.009 \\
& 30 & 2 & $8.59 \times 10^{9}$ & $9.68 \times 10^{-10}$ & 71561 & $9.67 \times 10^{-10}$ & 7861 & 0.054 & $9.67 \times 10^{-10}$ & 0.0960 & 0.054 & $9.67 \times 10^{-10}$ & 0.0006 & 0.054 \\
& 31 & 1 & $4.29 \times 10^{9}$ & $1.49 \times 10^{-9}$ & 143848 & $1.49 \times 10^{-9}$ & 15852 & 0.070 & $1.49 \times 10^{-9}$ & 0.0996 & 0.070 & $1.49 \times 10^{-9}$ & 0.0006 & 0.070 \\
\midrule
\textit{Int}$^{*}$
& 32 & 0 & $4.29 \times 10^{9}$ & $1.45 \times 10^{-9}$ & 188457 & $1.45 \times 10^{-9}$ & 24064 & 0.072 & $1.45 \times 10^{-9}$ & 0.0800 & 0.072 & $1.45 \times 10^{-9}$ & 0.0005 & 0.072 \\
\bottomrule
\multicolumn{15}{l}{\large $^{*}$Configurations for $m \leq 31$ were run in a single batch job separate from $m=32$, $e=0$; times between them are not directly comparable.} \\

\end{tabular}
}
\end{table*}


\vspace{-2mm}

\section{Conclusion}
\label{sec:conclusion}
\vspace{-1mm}

This paper introduced Weighted Integer (WINT), a configurable number representation format where numbers are expressed in the form $M \times 2^E$ with user-selectable mantissa ($m$) and exponent ($e$) bit allocations. We characterize WINT error using Mean Relative Error (MRE), including a harmonic approximation (HA) method that leverages the Euler-Maclaurin expansion to compute MRE in O(1) time complexity, achieving significant speedups over the integration method (IM) while maintaining accuracy within 0.2\% for the configurations presented. Our experiments across 8 to 32-bit configurations demonstrate that allocating 2 exponent bits consistently doubles the representable range while reducing MRE by 12--33\% compared to the integer baseline for bit widths of 12 and above. Allocating 3 exponent bits extends range by $16\times$ while reducing MRE by 15--50\% for bit widths of 16 and above. WINT's user-selectable bit allocation allows engineers to choose the specific mantissa-exponent configuration that best matches their application's precision and range requirements within a fixed bit budget.


\nocite{*}

{\footnotesize
\bibliographystyle{IEEEtran}

\bibliography{references}

@ARTICLE{ieee754_citation,
  author={},
  journal={IEEE Std 754-2019 (Revision of IEEE 754-2008)}, 
  title={{IEEE Standard for Floating-Point Arithmetic}}, 
  year={2019},
  volume={},
  number={},
  pages={1-84},
  doi={10.1109/IEEESTD.2019.8766229}
}

@Article{embedded_quantization,
  author = {Novac, Pierre-Emmanuel and Boukli Hacene, Ghouthi and Pegatoquet, Alain and Miramond, Beno\^{i}t and Gripon, Vincent},
  title = {{Quantization and Deployment of Deep Neural Networks on Microcontrollers}},
  journal = {Sensors},
  volume = {21},
  year = {2021},
  number = {9},
  article-number = {2984},
  url = {https://www.mdpi.com/1424-8220/21/9/2984},
  issn = {1424-8220},
  doi = {10.3390/s21092984}
}

@misc{fp8_citation,
  title={{FP8 Formats for Deep Learning}}, 
  author={Paulius Micikevicius and Dusan Stosic and Neil Burgess and Marius Cornea and Pradeep Dubey and Richard Grisenthwaite and Sangwon Ha and Alexander Heinecke and Patrick Judd and John Kamalu and Naveen Mellempudi and Stuart Oberman and Mohammad Shoeybi and Michael Siu and Hao Wu},
  year={2022},
  eprint={2209.05433},
  archivePrefix={arXiv},
  primaryClass={cs.LG},
  url={https://arxiv.org/abs/2209.05433}
}

@misc{bfp_citation,
  title={{Training DNNs with Hybrid Block Floating Point}}, 
  author={Mario Drumond and Tao Lin and Martin Jaggi and Babak Falsafi},
  year={2018},
  eprint={1804.01526},
  archivePrefix={arXiv},
  primaryClass={cs.LG},
  url={https://arxiv.org/abs/1804.01526}
}

@misc{mx_citation,
  title={{Microscaling Data Formats for Deep Learning}}, 
  author={Bita Darvish Rouhani and Ritchie Zhao and Ankit More and Mathew Hall and Alireza Khodamoradi and Summer Deng and Dhruv Choudhary and Marius Cornea and Eric Dellinger and Kristof Denolf and Stosic Dusan and Venmugil Elango and Maximilian Golub and Alexander Heinecke and Phil James-Roxby and Dharmesh Jani and Gaurav Kolhe and Martin Langhammer and Ada Li and Levi Melnick and Maral Mesmakhosroshahi and Andres Rodriguez and Michael Schulte and Rasoul Shafipour and Lei Shao and Michael Siu and Pradeep Dubey and Paulius Micikevicius and Maxim Naumov and Colin Verrilli and Ralph Wittig and Doug Burger and Eric Chung},
  year={2023},
  eprint={2310.10537},
  archivePrefix={arXiv},
  primaryClass={cs.LG},
  url={https://arxiv.org/abs/2310.10537}
}

@misc{flexpoint_citation,
  title={{Flexpoint: An Adaptive Numerical Format for Efficient Training of Deep Neural Networks}}, 
  author={Urs K\"{o}ster and Tristan J. Webb and Xin Wang and Marcel Nassar and Arjun K. Bansal and William H. Constable and O\u{g}uz H. Elibol and Scott Gray and Stewart Hall and Luke Hornof and Amir Khosrowshahi and Carey Kloss and Ruby J. Pai and Naveen Rao},
  year={2017},
  eprint={1711.02213},
  archivePrefix={arXiv},
  primaryClass={cs.LG},
  url={https://arxiv.org/abs/1711.02213}
}

@misc{ocp_mx_spec,
  author       = {{Open Compute Project}},
  title        = {{{OCP} Microscaling Formats ({MX}) Specification, Version 1.0}},
  year         = {2023},
  month        = {September},
  howpublished = {\url{https://www.opencompute.org/documents/ocp-microscaling-formats-mx-v1-0-spec-final-pdf}}
}

@article{posit_citation,
  author = {Gustafson, John L. and Yonemoto, Isaac},
  title = {{Beating Floating Point at its Own Game: Posit Arithmetic}},
  year = {2017},
  publisher = {South Ural State University},
  volume = {4},
  number = {2},
  issn = {2409-6008},
  url = {https://doi.org/10.14529/jsfi170206},
  doi = {10.14529/jsfi170206},
  journal = {Supercomputing Frontiers and Innovations},
  month = jun,
  pages = {71--86}
}

@misc{number_systems_survey,
  title={{Number Systems for Deep Neural Network Architectures: A Survey}}, 
  author={Ghada Alsuhli and Vasileios Sakellariou and Hani Saleh and Mahmoud Al-Qutayri and Baker Mohammad and Thanos Stouraitis},
  year={2023},
  eprint={2307.05035},
  archivePrefix={arXiv},
  primaryClass={cs.NE},
  url={https://arxiv.org/abs/2307.05035}
}

@Article{rns_isupov,
  author = {Isupov, Konstantin},
  title = {{High-Performance Computation in Residue Number System Using Floating-Point Arithmetic}},
  journal = {Computation},
  volume = {9},
  year = {2021},
  number = {2},
  article-number = {9},
  url = {https://www.mdpi.com/2079-3197/9/2/9},
  issn = {2079-3197},
  doi = {10.3390/computation9020009}
}

@article{euler_maclaurin_citation,
  title={{Euler's constant: Euler's work and modern developments}},
  volume={50},
  issn={1088-9485},
  url={http://dx.doi.org/10.1090/S0273-0979-2013-01423-X},
  doi={10.1090/s0273-0979-2013-01423-x},
  number={4},
  journal={Bulletin of the American Mathematical Society},
  publisher={American Mathematical Society (AMS)},
  author={Lagarias, Jeffrey C.},
  year={2013},
  month=jul,
  pages={527--628}
}

@manual{mpmath_citation,
  author     = {Fredrik Johansson and others},
  title      = {{mpmath: a {P}ython library for arbitrary-precision floating-point arithmetic (version 1.3.0)}},
  note       = {\url{https://mpmath.org/}},
  year       = {2023}
}
}

\end{document}